\documentclass[aps,10pt,twocolumn,prl]{revtex4-2}
\usepackage{amsmath}
\usepackage{amsfonts}
\usepackage{graphicx}

\begin{document}

\title{An $[\eta]$ Linear in $M$ Does Not Imply Rouse Dynamics}
\author{George D. J. Phillies}
\affiliation{Department of Physics, Worcester Polytechnic
Institute,Worcester, MA 01609}
\email{phillies@4liberty.net, 508-754-1859}

\begin{abstract}
Contrary to some expectations, an experimental finding for a polymer that the solution intrinsic viscosity $[\eta]$ or the melt viscosity is linear in the polymer molecular weight $M$ does not indicate that polymer dynamics are Rouselike.  Why? The other major polymer dynamic model, due to Kirkwood and Riseman [\emph{J. Chem.\ Phys.\ } \textbf{16}, 565-573 (1948)], leads in its free-draining form to a prediction $[\eta] \sim M$, even though the polymer motions in this model are totally unlike the polymer motions in the Rouse model. In the Rouse model, the chain motions are linear translation and internal ('Rouse') modes.  In the Kirkwood-Riseman model (and its free-draining form, derived here), the chain motions are translation and whole-body rotation.  The difference arises because Rouse's calculation implicitly refers only to chains subject to zero external shear force (And, as an aside, Rouse's construction of $[\eta]$ is invalid, because it concludes that there is viscous dissipation in a system that Rouse implicitly assumed to have no applied shear).
\end{abstract}

\maketitle

\section{Introduction}

Our objective is to correct a widespread misconception in the polymer dynamics literature.  It is often suggested that if the solution intrinsic viscosity or the melt viscosity is $\sim M$, then polymer dynamics are described at least approximately by the Rouse bead-spring model\cite{THrouse1953a}. This suggestion is here shown to be invalid, namely the Kirkwood-Risemann\cite{THkirkwood1948a} is also a bead-link model, bead motions in the Kirkwood-Riseman model are not at all Rouse-like, but in its free-draining form as obtained here the Kirkwood-Riseman description of bead motions also predicts $[\eta] \sim M$.

The Rouse and Kirkwood-Riseman calculations have one important similarity, namely they bifurcate. They both invoke bead-spring models, though in the Kirkwood-Riseman calculation fluctuations in the lengths and orientations of the 'springs' were as an approximation neglected.   In the first part of each calculation, the motions of the beads are calculated.  In the second part of each calculation, the models are used to calculate the intrinsic viscosity.

In the following, Section II presents in modern notation the relevant aspects of the Kirkwood-Riseman model\cite{THkirkwood1948a}.  Section III calculates the intrinsic viscosity $[\eta]$ for the free-draining form of the Kirkwood-Riseman model. As a Discussion, Section IV compares the Kirkwood-Riseman and Rouse descriptions of polymer dynamics, including points at which recent computer simulations\cite{THphillies2018a} clarify the differences between the models.

\section{The Kirkwood-Riseman Model}

 Two-thirds of a century ago, Kirkwood and Riseman advanced a treatment of polymer chains in solution, including solutions subject to a shear field. Their interests were the intrinsic viscosity $[\eta]$ and the self-diffusion coefficient $D_{s}$ of the polymers, and the dependences of these transport parameters on polymer molecular weight $M$. Their objective was to explain why, for long chains, $[\eta]$ and $D_{s}$ depend on $M$ sublinearly in $M$, i.e., $[\eta] \sim M^{a}$ with $a < 1$.  As an approach, they incorporated hydrodynamic interactions into a plausible description of a random-coil polymer in solution, leading to a prediction $[\eta] \rightarrow M^{0.5}$ as $M \rightarrow \infty$.

 In the Kirkwood-Riseman model, polymer coils were taken to be, on the average, spherically symmetric. The distances between pairs of beads, and between each bead and the polymer center of mass, were approximated as having fixed values calculated from an assumed Gaussian distribution of random walks for the coil contour.  Kirkwood and Riseman recognized that bead-bead distances fluctuate, but these fluctuations were taken to provide at most secondary corrections to calculated values for $D_{s}$ and $[\eta]$.

 Each bead was assumed to have a hydrodynamic drag coefficient $f$, so that the hydrodynamic force $\mathbf{F}_{\ell}$ on a bead $\ell$ was
\begin{equation}
    \mathbf{F}_{\ell} = f(\mathbf{u}_{\ell} - \mathbf{v}_{\ell})
    \label{eq:THhydroforce}
\end{equation}
Here $\mathbf{v}_{\ell}$ is the current velocity of bead $\ell$ and $\mathbf{u}_{\ell}$ is the velocity that the fluid would have had, at the location of bead $\ell$, if the bead were not present.  In the original paper, pairwise bead-bead hydrodynamic interactions were taken to be described by the Oseen tensor. In this paper, the free-draining model corresponding to the Kirkwood-Riseman model is considered, so there are no hydrodynamic interactions between pairs of beads.

Kirkwood and Riseman applied to their model chain a position-dependent fluid velocity flow $\mathbf{u}(\mathbf{r})$, and considered the response of a polymer molecule to the flow field. Their flow ran parallel to the $x$ axis, with a linear gradient $\dot{\epsilon}$ in the $y$ direction, so that the fluid velocity that would have been found at $\mathbf{r} = (x,y,z)$ if no bead were present at that location was
\begin{equation}
   \mathbf{u}(\mathbf{r}) = u_{0} \mathbf{\hat{i}} +  \dot{\epsilon} y \mathbf{\hat{i}},
   \label{eq:THvfluid}
\end{equation}
where $u_{0}$ is the x-component of the fluid velocity at $y=0$ and $\mathbf{\hat{i}}$ is the unit vector in the $x$-direction.

For the motion of individual polymer beads, Kirkwood and Riseman then partitioned the bead velocity $\mathbf{v}_{\ell}$ into a whole-chain translational component $\mathbf{V}$, a whole-chain rotational component $\mathbf{\Omega}$, and an internal component here denoted $\mathbf{e}_{\ell}$, namely
\begin{equation}
      \mathbf{v}_{\ell} = \mathbf{V} + \mathbf{\Omega} \times \mathbf{r}_{\ell} + \mathbf{e}_{\ell}.
      \label{eq:THvelocity}
\end{equation}
Here $\mathbf{r}_{\ell}$ is the position of bead $\ell$ relative to the chain's center. Equation \ref{eq:THvelocity} is effectively the definition of $\mathbf{e}_{\ell}$. The internal component $\mathbf{e}_{\ell}$ was dropped from further consideration because the model does not consider fluctuations in bead positions around their average locations.  The polymer coil was effectively treated as a rigid body.  $\mathbf{V}$ and $\mathbf{\Omega}$ were obtained from the observation that the model is a lower-frequency approximation, so the polymer's inertia is neglected.  It follows that the net force and net torque on the polymer chain as a whole, on the time scales of interest, average very nearly to zero.

For the net force on a polymer coil, one may write a sum over the $N$ beads of the chain as
\begin{equation}
    \sum_{\ell=1}^{N} \mathbf{F}_{\ell} = \sum_{\ell=1}^{N}   f(u_{0} \mathbf{\hat{i}} + y_{\ell} \dot{\epsilon} \mathbf{\hat{i}} - (\mathbf{V} + \mathbf{\Omega} \times \mathbf{r}_{\ell})).
    \label{eq:THnetforce1}
\end{equation}
The left-hand-side of this equation is zero by the no-net-force condition.  On the right hand side of the equation, terms in $y_{\ell}$ and $ \mathbf{r}_{\ell}$ sum to zero because the coil is taken to be spherically symmetric, leaving
\begin{equation}
     \mathbf{V} = u_{0} \mathbf{\hat{i}},
     \label{eq:THdrift velocity}
\end{equation}
where the center of mass of the polymer has been taken to be $y=0$, i.e., the polymer coil simply drifts with the average flow.

For the net torque on the polymer coil, one may write
\begin{equation}
    \sum_{\ell=1}^{N} \mathbf{r}_{\ell} \times \mathbf{F}_{\ell} = \sum_{\ell=1}^{N}  \mathbf{r}_{\ell} \times  f(u_{0} \mathbf{\hat{i}} + y_{\ell} \dot{\epsilon} \mathbf{\hat{i}} + (\mathbf{V}- \mathbf{\Omega} \times \mathbf{r}_{\ell})).
    \label{eq:THnettorque1}
\end{equation}
The left-hand-side of this equation is again zero.  From spherical symmetry, on expanding in Cartesian coordinates, terms in $\langle x_{\ell}^{2} \rangle$, $\langle y_{\ell}^{2} \rangle$, and $\langle x_{\ell}^{2} \rangle$ survive (these terms are equal), while cross terms such as $\langle x_{\ell} y_{\ell} \rangle$ average to zero, leading to
\begin{equation}
       \mathbf{\Omega} = - \frac{1}{2} \dot{\epsilon} \mathbf{\hat{k}}
       \label{eq:THnettorque}
\end{equation}
For positive shear $\dot{\epsilon}$,  the chain as seen from above rotates in the clockwise direction.  In the Kirkwood-Riseman model, polymer chains in shear translate to match the average velocity of the solvent within the chain; they also rotate in such a way that the average net torque on a chain vanishes.

Bead-bead hydrodynamic interactions, such as those described by the Oseen tensor, do not enter the above calculation, because bead-bead interactions are purely internal forces, and from Newton's Third Law purely internal forces cannot create a new force or a net torque on an object. Equations \ref{eq:THdrift velocity} and \ref{eq:THnettorque} are therefore equally correct for non-draining and for free-draining polymer chains.

\section{Intrinsic Viscosity of the Freely-Draining Rouse Model}

We now advance to calculate the intrinsic viscosity of the freely-draining Kirkwood-Riseman model.

Kirkwood and Riseman note two possible approaches to calculating the intrinsic viscosity, namely for their model one could calculate how a polymer molecule perturbs the flow at the fluid boundaries, or one could calculate how a polymer molecule contributes to dissipation.  Kirkwood and Riseman used the second of these approaches, as applied to a non-draining polymer chain in which the fluid flow around each bead and the corresponding power dissipation are altered by the hydrodynamic interactions between the beads, which must be treated in a self-consistent way.

For the free-draining model considered here, the calculation of the dissipation is straightforward.  The rate at which work is done on a polymer bead $\ell$ is $\mathbf{F}_{\ell} \cdot \mathbf{v}_{\ell}$, where $\mathbf{F}_{\ell}$ and $\mathbf{v}_{\ell}$ follow from the previous section.  Dissipation arises because the polymer beads move in accord with eq.\ \ref{eq:THvelocity}, so they cannot all move at the same velocity as the neighboring fluid.  For example, as a result of the $\mathbf{\Omega} \times \mathbf{r}_{\ell}$ cross-product, most beads will have a velocity component parallel to the $y$-axis, even though there is no fluid motion in that direction.  The power passing via hydrodynamic friction through all the chains in the system is then
\begin{equation}
     P = \langle \sum_{m=1}^{N_{c}} \sum_{\ell=1}^{N} \mathbf{F}_{\ell} \cdot \mathbf{v}_{\ell} \rangle
      \label{eq:THpower1}
\end{equation}
Here $m$ is a sum over the system's $N_{c}$ chains and $\ell$ is a sum over the $N$ beads of a single chain.  $N_{c} N$ is the number of beads in the system; it is independent of the polymer's molecular weight. $P$ may be rewritten as
\begin{equation}
     P = \langle N_{c} N f (u_{o} \mathbf{\hat{i}} + \dot{\epsilon} y_{l} \mathbf{\hat{i}} - \mathbf{V} -\mathbf{\Omega} \times \mathbf{r}_{\ell}) \cdot (\mathbf{V} -\mathbf{\Omega} \times \rangle \mathbf{r}_{\ell})
     \label{eq:THpower2}
\end{equation}

The force term has components that cancel.  Equation \ref{eq:THpower2} is readily evaluated by replacing all vectors with their cartesian components, leading to
\begin{equation}
   P=0.
   \label{eq:THpower3}
\end{equation}
$P$ vanishes because it has two parts that cancel: The shear field is transferring energy to the polymer chain; frictional resistance to chain rotation removes energy from the polymer chain.  Because we are in steady state, the rotational energy of the chain does not change as time advances, so the two components must cancel.  The dissipation, the second term, is the polymer contribution to the viscosity.  Noting that the radius of gyration satisfies $R_{g}^{2} = \langle x_{\ell}^{2} \rangle + \langle y_{\ell}^{2} \rangle  + \langle z_{\ell}^{2} \rangle$, the three terms on the right being equal to each other, the dissipation by the polymer chains becomes
\begin{equation}
   P = - [N_{c} N] \eta_{0} [\frac{f}{\eta_{0}} \frac{R_{g}^{2}}{3}] \frac{\dot{\epsilon}^{2}}{2}.
   \label{eq:THpower4}
\end{equation}
Here the term in the first pair of square brackets counts the number of beads in the system.  By convention a factor of the solvent viscosity has been factored out. The second term in square brackets is, up to constants, the intrinsic viscosity $[\eta]$.  The power dissipation is seen to be proportional to the square of the shear rate $\dot{\epsilon}$.  

Observe that the free-draining Kirkwood-Riseman form for $P$ does not include any internal mode relaxation times.  In this expression, only $R_{g}^{2}$ depends on the polymer molecular weight, leading to
\begin{equation}
    [\eta] \sim M.
    \label{eq:THpower5}
\end{equation}

The free-draining Kirkwood-Riseman model thus agrees with the Rouse model in its prediction that the intrinsic viscosity depends linearly on the polymer molecular weight. Claims that an observation $[\eta] \sim M$ for a polymer melt supports the presence of Rouse-like polymer motion are therefore incorrect, because the entirely-contrary free-draining Kirkwood-Riseman description makes the same prediction.

\section{Discussion}

In the Kirkwood-Riseman model, translation is described by 3 degrees of freedom; rotation is also described by three degrees of freedom.  An $N$-particle chain is described by $3N$ coordinates, $3N$ degrees of freedom, so there are $3N-6$ remaining coordinates to describe a Kirkwood-Riseman chain's internal motions.  However, within this model, the internal motions in which the chain's beads move with respect to each other are taken not to contribute substantially to the intrinsic viscosity.

We may immediately contrast this description of the modes of a Kirkwood-Riseman model chain with a description of the modes of a Rouse chain.  In the Rouse model, chain motions in the $x$, $y$, and $z$ directions are independent from each other, so the Rouse model can be solved by treating motions parallel to any one of the three coordinates.  Of course, a Rouse chain is three-dimensional, so the solution is replicated for each coordinate axis. For each coordinate axis, a Rouse chain has $N$ normal coordinates.  One of these corresponds to uniform translation of all polymer beads parallel to the relevant coordinate axis.  The other $N-1$ normal coordinates correspond to internal modes in which polymer beads move with respect to each other.  In total, a Rouse chain thus has three translational modes and $3N-3$ internal modes, for a total of $3N$ modes, leaving no remaining modes to be assigned to rotation.

These Kirkwood Riseman and Rouse/Zimm models for polymer motion are manifestly contradictory in their descriptions of how polymers move in solution.  In one model, the chain is claimed to translate and rotate, internal motions being negligible.  In the other model, the chain is claimed to translate and have internal motions, rotation being negligible.

The notion that Rouse chains do not rotate is asserted by Rouse.  Rouse\cite{THrouse1953a} specifies that a polymer coil under shear does not rotate, namely (his paper, p. 1274, column 2) "\emph{...since the velocity of the liquid has a nonvanishing component only in the $x$ direction, the components $(\dot{y}_{j})_{\alpha}$ and $(\dot{z}_{j})_{\alpha}$ are zero.}"  $(\dot{y}_{j})_{\alpha}$ and $(\dot{z}_{j})_{\alpha}$ are the velocities of bead $j$ in the $y$ and $z$ directions due to the shear. However, if the chain is to be rotating, either $(\dot{y}_{j})_{\alpha}$ or $(\dot{z}_{j})_{\alpha}$  must be non-zero, so, according to Rouse, a shear field leads to chain distortion but not chair rotation.

Rouse also asserts in his paper (p. 1274, column 2, top) that '\emph{...an atom at the junction between two submolecules...}' (springs) moves '\emph{...with a velocity equal to that of the surrounding liquid...}' except for Brownian motion, because, according to Rouse, otherwise there would be motion of the solvent relative to the polymer chain, leading to energy dissipation. The fluid only moves in the $\mathbf{\hat{i}}$ direction, so if the beads only move with the liquid, then they can only be moving parallel to the $x$-axis.  Under the influence of shear, a Rouse polymer may be very loosely imagined as distorting from a rectangular to a parallelogram form, with no tendency for rotation to occur.

It may appear odd that Rouse's chains do not rotate.  One might also wonder how, if the beads of a Rouse chain simply move with the solvent flow so that there is no dissipation, Rouse's chains can be contributing to the viscosity.  A trivial modification of Rouse's original equations of motion eliminates these issues.  However, with the modification Rouse's solutions to his model are invalid.  

Rouse\cite{THrouse1953a} wrote for the equation of motion of a typical polymer bead (end beads are special cases)
\begin{equation}
    f \frac{d \mathbf{{r}}_{i}}{dt} =  k (\mathbf{{r}}_{i-1}  +  \mathbf{{r}}_{i+1} - 2 \mathbf{{r}}_{i} ),
     \label{eq:THrousemotion}
\end{equation}
$k$ being an effective spring constant.  In this equation there is no term representing the force $f \dot{\epsilon} y_{i}\mathbf{\hat{i}}$ due to the shear field. 
If a shear field is added to equation \ref{eq:THrousemotion}, the equation of motion of a typical bead would instead read
\begin{equation}
    f \frac{d \mathbf{{r}}_{i}}{dt} =  k (\mathbf{{r}}_{i-1}  +  \mathbf{{r}}_{i+1} - 2 \mathbf{{r}}_{i} ) + f \dot{\epsilon} y_{i}\mathbf{\hat{i}}.
     \label{eq:THtruemotion}
\end{equation}
The reason that Rouse's chains do not rotate is immediately obvious.   Equation \ref{eq:THrousemotion} is simply the special case of eq.\ \ref{eq:THtruemotion} for the case $\dot{\epsilon} = 0$.  If $\dot{\epsilon} = 0$, there is no rotation because the solvent is quiescent; there is no shear field to drive rotation. The beads move with the local fluid flow, meaning that they are all stationary with respect to the solvent.  There is then no dissipation. Rouse's calculation of viscous dissipation due to a polymer chain, within his model, cannot be correct, because within his model the shear rate $\dot{\epsilon}$ is zero, so there is no polymeric contribution to the viscosity. Nonetheless, Rouse constructed from his model an intrinsic viscosity satisfying $[\eta] \sim M$.

How do bead-spring polymers move when shear is applied?  Rouse coordinates, being a special case of the discrete Fourier transform, remain valid.  However, as shown by recent single-chain Brownian dynamics calculations\cite{THphillies2018a}, when Rouse chains are subject to shear:
      \begin{itemize}

      \item The polymer chain does indeed rotate.

      \item Some Rouse modes become cross-correlated.

      \item The amplitudes of the Rouse modes depend on the shear rate.

      \item The relaxation rates of Rouse modes depend on the shear rate.

      \end{itemize}

These properties of Rouse chains under shear are, unsurprisingly, not the same as the properties found by Rouse for bead-spring chains in the absence of an applied shear field. More important, they are not the properties oft assumed for Rouse chains under shear. Bead-spring chains under shear have three translational modes, three rotational modes, and therefore can have only $3N-6$ internal modes, not the $3N-3$ internal modes of the Rouse model.

Finally, free-draining Kirkwood-Riseman chains, which translate and rotate but have no internal modes, cannot be distinguished from hypothetical chains that satisfy Rouse's construction of an intrinsic polymer viscosity, because both types of chain have $[\eta] \sim M^{a}$ for $a=1$.  Correspondingly, an experimental finding $[\eta] \sim M$ is not evidence for Rouse-like polymer motion because a polymer chain following free-draining Kirkwood-Riseman dynamics follows $[\eta] \sim M$.

\end{document}